\documentclass{ifacconf_amx}

\usepackage{graphicx}      
\usepackage{svg} 	 	   
\usepackage{amsmath,amssymb,amsfonts} 
\usepackage{cite}          
\usepackage{float}

\usepackage{siunitx}       
\usepackage{fancyhdr}      

\usepackage{lipsum}        

\usepackage{tabularx}	   
\usepackage{booktabs}	   
\usepackage{xcolor}        
\usepackage{multirow}	   

\usepackage[scaled]{helvet}
\usepackage[T1]{fontenc}

\newcommand{\chair}{Technical University of Munich\\Professorship of Agrimechatronics}
\newcommand{\amxadress}{Technical University of Munich, Germany; Professorship of Agrimechatronics; \\ Munich Institute of Robotics and Machine Intelligence (MIRMI)}

\makeatletter

\newcommand{\AuthorList}{}
\let\AuthorHeader\empty

\newcommand{\AddAuthor}[3][]{%

	\g@addto@macro\AuthorList{%
		\author[#3]{#2}%
	}

	\def\@tempname{#1}
	\ifx\@tempname\@empty
		\def\@tempname{#2} 
	\fi

	\ifx\AuthorHeader\empty
		\xdef\AuthorHeader{\@tempname}%
	\else
		\xdef\AuthorHeader{\AuthorHeader, \@tempname}%
	\fi
}

\let\old@ssect\@ssect 
\makeatother

\usepackage{hyperref}      
\hypersetup{hidelinks}
\usepackage{cleveref}      

\makeatletter
\def\@ssect#1#2#3#4#5#6{%
	\NR@gettitle{#6}
	\old@ssect{#1}{#2}{#3}{#4}{#5}{#6}
}
\makeatother

\usepackage{silence}
\usepackage{placeins}
\usepackage{subcaption}

\newcommand{\ISBN}{978-3-911430-17-3}
\newcommand{\PaperTitle}{Digital Twin Modeling of a Highly Automated Agricultural Tractor}
\newcommand{\PaperDate}{22 July 2026}
\newcommand{\PaperKeywords}{Digital Twin, Agricultural Tractor, Kinematics, CAN Communication}

\AddAuthor[Hallman C.]{Clay Hallman}{AMX}
\AddAuthor[Pindl L.]{Lukas Pindl}{AMX}
\AddAuthor[Oksanen T.]{Timo Oksanen}{AMX}

\hypersetup{
    pdfinfo={
        Title={\PaperTitle},
        Author={\AuthorHeader},
        Keywords={\PaperKeywords},
        Publisher={Technical University of Munich},
        ISBN={\ISBN}
    }
}

\begin{document}
\begin{frontmatter}

   \title{\PaperTitle}

   \AuthorList

   \address[AMX]{\amxadress}

   \begin{abstract}
      In efforts to increase research efficiency and availability, a digital twin of our research tractor (AMX G-trac) is created, focusing especially on the CAN communication for data reading and actuation command following the ISOBUS protocol. Mevea Simulation Software is utilized as the foundation, providing the kinematic model and visuals, while Python is used to read and write CAN messages over a Kvaser CanKing virtual CAN channel. Various     performance tests involving straight line and turning behavior are performed in both the     digital twin simulation and in the real world to measure similarity. Results indicate that
      the Mevea model behaves very comparable in its lateral dynamics, often within 5-10
      percent, but requires better data to fully capture the longitudinal aspects like     acceleration. The final model described in this paper sets the table for a second iteration
      to include more tractor functions such as hydraulics and tractor-implement dynamics.
   \end{abstract}


\end{frontmatter}

\section{Introduction}

In the automotive and aerospace sectors, digital twins have already been established as
a necessary part of development. While some work has been conducted \cite{foldager2021agricultural}, much less researched is the topic of digital
twins for agricultural tractors \cite{Purcell2023AgricultureDT}. While tractor manufacturers make up a much smaller
portion of vehicle OEMs than passenger car brands, both in quantity and revenue, the
realizable benefits from digital twin-based innovations for tractors should not be
underestimated. By improving system models of tractors through digital twin processes
e.g. what-if scenarios, only then can the correct design decisions be made \cite{Zhizdyuk2025MachineryDT,soderberg2017digitaltwin,piromalis2022automotive}.

\begin{figure}
   \centering
   \includegraphics[width=0.95\linewidth]{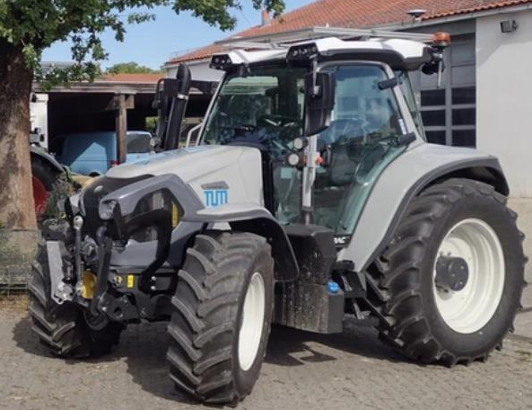}
   \caption{AMX G-trac research tractor}
   \label{fig:gtrac}
\end{figure}

Our research tractor called AMX G-trac is a platform for studying subsystems of autonomy and future functions of agricultural tractor (see Figure \ref{fig:gtrac}). The tractor is based on Lindner Lintrac 130. Some examples of applied research on this tractor include tasks such as coverage path planning and other autonomous driving modules, as well as this research into digital twins for tractors. By looking into this topic, research can become more efficient. The access to a real-world accurate simulation provides the ability to pre-test concepts indoors, removing dependencies on space, weather or general access to the tractor. Furthermore, potentially dangerous questions involving actuation for example can first be investigated and improved before applying approaches to the real-world tractor, where safety and costs are highly relevant.

\section{State of the Art}

There is a plethora of modern solutions to create a digital twin. Companies like Siemens
and Ansys offer powerful programs but with specialties that lie outside of this project. Specifically, agricultural tractors depend more on fuel usage, maintenance and downtime costs than crash safety for example in passenger vehicles \cite{liu2021drivingstate}.
For the task of developing a physically accurate multi-body dynamic simulation with
special consideration of heavy machinery and soil mechanics, the Mevea Simulator software is
currently a state-of-the-art solution.
Mevea offers the most advanced physics engine available alongside real-time simulation
\cite{Mevea2026Website}. Most of their technology is tailored towards port, construction or mining equipment, often with the end goal of operator training in mind. Although there have
been a few others to consider Mevea for agricultural tractors \cite{Jaiswal2026Configurator} or other industrial equipment \cite{panwar2024excavator,pyne2021heavyduty}, it is still a very much untapped opportunity. Mevea’s expertise in off-road machinery makes it a perfect fit when trying to accurately incorporate all the real-world physical considerations into a digital twin.
Outside of its dynamic modeling capabilities, it also specializes in hydraulic simulation,
which is responsible for many functions within the tractor. Albeit outside the scope of
this project, tractor-implement interaction could therefore be accurately investigated, as
well as brake functionality.
Mevea also has several different options for CAN network communication. PCAN
compatibility allows for semi-direct plug and play assuming physical CAN channels are
available. If virtual CAN channels are used (as in this case), Mevea also offers simple
ways to extract the component data in real time, thanks to its Python environment.

\section{Materials}

As briefly mentioned earlier, the focus points for this project are primarily the basic
kinematic functions of the tractor, and the ability to simulate CAN message-controlled
actuation to investigate the vehicle's reactionary behavior.

\subsection{Software}

Seeing that a digital twin is a software-heavy undertaking, the majority of materials that
were needed to complete this research also lie in various software applications.
First, the backbone of this topic, where the digital twin lives and exists is the Mevea
Simulation Software. Here, the full physical model is built and the multibody dynamic
simulation runs in real time. Here, all aspects of the tractor's kinematics can be
considered, such as driving, steering, suspension or the tire-soil interaction. Tractor
implement dynamics can also be included but were left out in this project. Plus, Mevea
has the capability to control simulation inputs via Python script, which allows for a fully
virtual CAN based communication for actuation.
Kvaser CanKing is used to set up the virtual CAN channel on which messages are both
sent and received, while various Python scripts control the reading and writing of the
CAN messages.
To create a visually accurate tractor body, Giants Editor was utilized to import the
Farming Simulator 2025 Lintrac130 model into Blender. There the model can be
exported to a suitable file format for SolidWorks. This gives the possibility of finding
rough inertia tensors and component masses as starting points for stabilizing the
kinematic performance of the tractor. Here, the axle and body coordinate systems can
also be defined to provide simple graphical importing into Mevea.

\subsection{Hardware}

As the name implies, a digital twin involves a physical vehicle as well. In this case, the physical appearance of the AMX G-trac is mostly following original Lindner Lintrac 130 tractor (see Figure \ref{fig:gtrac}). This 136 hp diesel engine tractor includes four-wheel steering capabilities and a power-split CVT transmission. The AMX G-trac has been further modified to allow for individual control over each axle, enabling the classic tight turning but also crab-steering.

\section{Methods}

\subsection{Modeling}

The modeling process is split into two different sections - the physical multi-body
dynamics modeling and the necessary steps required to provide a graphical
representation.

\begin{figure}
   \centering
   \includegraphics[width=0.95\linewidth]{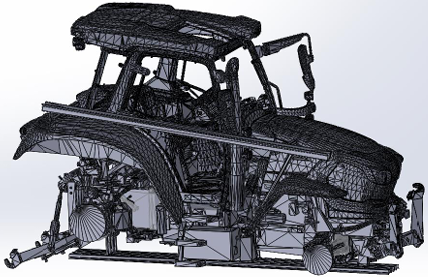}
   \caption{\textit{Farming Simulator 25} Tractor Model in SolidWorks}
   \label{fig:3DTrac}
\end{figure}

Much of the heavy lifting for the actual kinematic model was done by getting the
necessary parameters. This was accomplished through a few different methods. First, total weight, tire sizes and weights and drivetrain values were taken from the relevant manufacturers' specification sheets. For the kinematic model however, one total weight is not sufficient, as each kinematically responsible body needs to be individually defined. This means the front and rear axles, and the remaining bodies (grouped into one cab) each need their own weights, inertia tensors and relevant distances between one another. For this purpose, the openly available 3D model from \textit{Farming Simulator 25} was used (Figure \ref{fig:3DTrac}).

By exporting this model into CAD software (in this case SolidWorks), the relevant distances between the necessary body coordinate systems can immediately be measured (making sure to cross-reference some basic CAD model measurements with the real tractor for verification). This is of course also possible to do on the AMX G-Trac itself, however the CAD software allows for much easier access to components in the mid-section of the tractor. Since weights and inertia tensors for each individual body are not provided by manufacturers, the surface mesh from the \textit{Farming Simulator 25} export was then split into the relevant bodies. For the front and rear axle, assumptions were then made as to material and part density based on common axle weight values. Finally, using the \textit{Mass Properties} tool in SolidWorks, the axle weights and inertia tensors could be estimated. The imported mesh body of the tractor meant that there were thousands of surfaces just for the cab. Mainly due to time and compute constraints, a simplified model of the tractor
cab was created from scratch in SolidWorks. Taking the measured total weight of the
tractor and subtracting the axle weights from the mass properties tool, a tractor body
weight, comprising of the cab, transmission, engine and other attached components can
be found. Then, based on moment equilibrium around the tire contact points, the length
wise COG position for the whole tractor (with axles) can be calculated.

\begin{figure}
   \centering
   \includegraphics[width=0.95\linewidth]{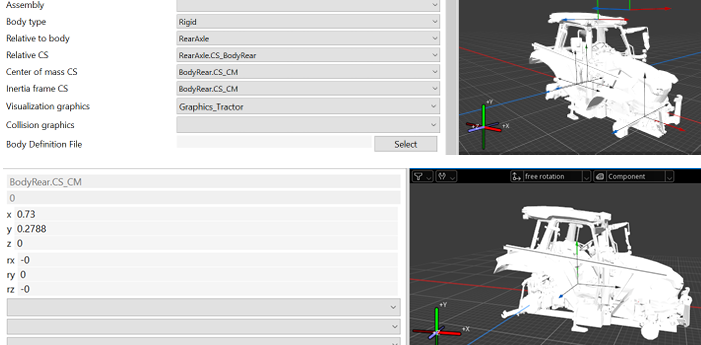}
   \caption{Creation of Kinematic Chain}
   \label{fig:kin_chain}
\end{figure}

For implementation into Mevea, the important steps to take care of when inserting the values into
the relative body fields are then to select from which body each is referred to, and the
location of the coordinate system for that body relative to its reference body. An example is shown in Figure \ref{fig:kin_chain}.

\begin{figure}[H]
   \centering
   \includegraphics[width=0.95\linewidth]{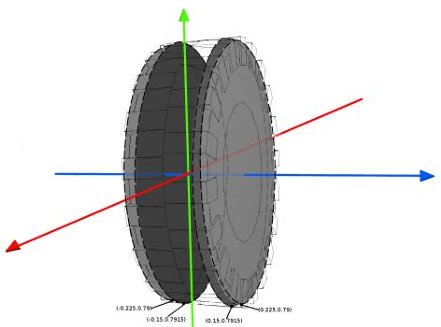}
   \caption{Tire Spline Values.\cite{MeveaModellerTutorials2026}}
   \label{fig:tire_spline}
\end{figure}

Typically, only one axle needs to be given a coordinate system in order to be controlled for steering. However, a unique feature of the research tractor (AMX G-trac) is its dual axle steering. Due to this fact, each axle needs to be modeled individually in order to allow for independent steering.
To accomplish dual axle steering, a couple of different options are available. A detailed
modeling of the steering geometries and knuckles is possible, but for simplicity, it was
decided to rather model the axles as rotating themselves with a steering input and relate the axle bodies to the tractor body using realistic constraints like revolute joints at the
axle-tractor body mounting points. This works kinematically but could lead to
unrealistic visualizations. The solution to this is presented in the next section.
Constraints are what allow the different bodies to realistically move in relation to one
another. In the context of constraints, apart from the two revolute joints at the axle
connections, the only other necessary piece is to constrain the rear axle to the ground.
This then enables the realistic motion of the vehicle across the surface without modeling
suspension, and all other bodies move through the defined joints relative to its previous
body in the kinematic chain. Suspension can also be modeled as a revolute joint around
the longitudinal axis rather than the vertical axis, but when coupled with steering a new
joint type must be defined to allow for both directions of rotation.
For tire modeling, the LuGre method is applied, requiring lateral and longitudinal
friction values for the tires as well as spring and damping constants. These values are taken to be the same as the Mevea default model values. A tire spline (Figure \ref{fig:tire_spline}) is also given to create the shape of the tires through its radius and various coordinates.

\begin{figure}
   \centering
   \includegraphics[width=0.95\linewidth]{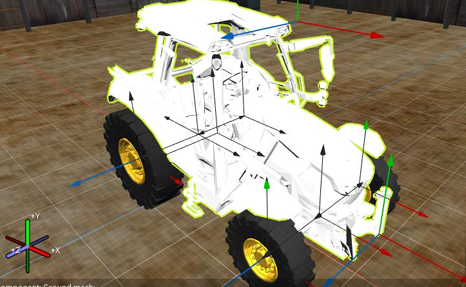}
   \caption{Mevea Model with all Coordinate Systems.}
   \label{fig:tractorCS}
\end{figure}

The final model with all its various coordinate systems then looks like that in Figure \ref{fig:tractorCS}

\subsection{Visualization}

While the model in Mevea will run smoothly without any visual bodies, it is often more
helpful and intuitive to include a 3D representation rather than solely relying on graphs
and numeric outputs. In Mevea this is accomplished by assigning 3D files such as .3ds
or .stl to the corresponding kinematic body. The graphics will in no way affect the
dynamic simulation but will simply provide visual feedback as to the simulation’s
behavior.
The graphics used in this project are the full tractor and the tires. Due to the way in
which the steering is implemented (explained in the previous section), if the axles were
to be their own graphic bodies, they would simply just turn underneath the tractor body.
However, by giving the full tractor 3D model to the tractor body graphic, only the tires
will turn when steering, providing a much more realistic visualization while still
allowing for a reasonable steering mechanic.
The environment in which the tractor drives has already been provided from Mevea standard models.

\subsection{CAN bus and communication with Mevea simulation}

To set up the virtual CAN bus, both Python and Kvaser CanKing were utilized. Kvaser
CanKing was responsible for setting up the virtual CAN channel which allows data to
be sent as defined in the CAN network protocol i.e. a specific baud rate and messages
consisting of various bytes for IDs and message data etc. It can also be used for bus
traffic monitoring.
Sending messages on this virtual channel was done via the proprietary library for CAN communication. The CAN library serves as the foundation for sending and receiving CAN messages, while the custom library processes incoming messages and formats outgoing messages according
to predefined message handlers.

\begin{figure}[H]
   \centering
   \includegraphics[width=0.95\linewidth]{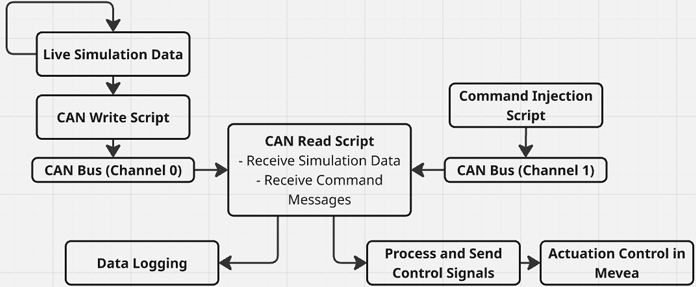}
   \caption{Data Flowchart}
   \label{fig:dataflow}
\end{figure}

While Mevea offers various ways to connect CAN channels to the simulation, for a fully
virtual channel, a simple way was to communicate through text files. Mevea does
support PCAN networking for example, and other predefined data transfer methods with
the help of its Nets Configuration Tool, however a professional version of PCAN is
needed to create a virtual bus, and due to the specific license, this tool was unavailable.
Within a Mevea Python script, any desired variables can be written to a text file. Each
file then stores the float value of its respective simulation variable, where it can be read
by the first CAN Python script running outside of the simulation. This script then writes
the CAN message onto the virtual channel where it can be read by another external script
for purposes of data recording and visualization. Through one final Python script,
desired actuation inputs can then be specified over CAN, simulating the real-world
process of autonomous control. Here, text files also have the advantage that while testing
the Mevea model, values can just be quickly updated by modifying the text file directly
rather than doing it all through CAN. Finally, when the CAN message is recognized as
a command, it writes that value to an input text file which is then read by the input script
running each actuator in Mevea. A full example flow overview of simulation data with
command injections is provided (Figure \ref{fig:dataflow}).

\section{Results}
\begin{table*}
   \centering
   \caption{Turn Test Results}
   \label{tab:turning_results}

   \newcolumntype{Y}{>{\centering\arraybackslash}X}
   \begin{tabularx}{\textwidth}{Y Y Y Y Y Y Y}
      \toprule
      Test                                    &
      Mean True Front Axle Steering Angle (°) &
      Mean Sim Front Axle Steering Angle (°)  &
      \% Difference Front Axle                &
      Mean True Rear Axle Steering Angle (°)  &
      Mean Sim Rear Axle Steering Angle (°)   &
      \% Difference Rear Axle                                                                  \\
      \midrule

      Turn Test 10/10                         & 10.77 & 10.87 & 0.95  & 9.92   & 10.86 & 9.5   \\
      Turn Test 20/0                          & 19.65 & 20.25 & 3.08  & 0.183  & 0.181 & 1.5   \\
      Turn Test 35/0                          & 32.48 & 35.97 & 10.74 & 0.15   & 0.38  & 160   \\
      Turn Test 20/-6                         & 19.92 & 20.23 & 1.57  & -5.76  & -4.06 & 29    \\
      Turn Test 35/-12                        & 32.30 & 35.87 & 11.00 & -11.82 & -8.00 & 36.65 \\
      \bottomrule
   \end{tabularx}

\end{table*}

For the results of the turn tests, Table \ref{tab:turning_results} shows the relative averages and percent
difference between the average values of each test’s set target steering angle values over
the time window.  Despite there being a slight buffer area of 1 degree in the steering
angle controller implemented in Python, while observing the simulation runs in real
time, it becomes apparent that the majority of the errors in the steering axles from the
simulation actually stems from the data transfer process and not the physics itself.
Although the axle behaves stable within the Mevea simulation, the solver timesteps
seemed to be too quick for the text file method of data transfer to always complete the
full steering angle calculations correctly, leading to the spikes seen in the
charts of the non-crab steering tests (Figure \ref{fig:appendix1}). This is further supported by the front axle spikes
only occurring once the full turn is made, where large angle values get reset to small
ones. It should be noted that the time scales are different due to tractor speeds as well as
processing power of the computer running the simulation, however each tractor made
roughly two complete turns.

Further evidence of this being caused by processing speeds is the fact that the turning
behavior of both the real tractor and the twin are the same. That is to say, these spikes
do not actually show up in the heading data, because as mentioned before, the simulation
runs smoothly and the steering angles are calculated subsequently. The curves in Figure \ref{fig:appendix2}
have been shifted and time scaled to be able to view each individually, as the focus is on
the similar behavior and the CAN data recording, rather than the speed in which the
tractors turn.

For the crab steering (10/10) turning test, most of the noise came from the front axle of
the real-world tractor, although it is still well within an acceptable range.

\begin{figure}
   \centering
   \includegraphics[width=0.95\linewidth]{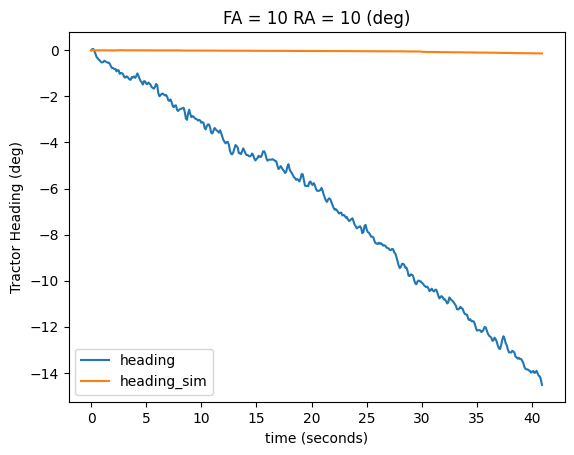}
   \caption{Crab Steering Tractor Heading over Time}
   \label{fig:crabheading}
\end{figure}

The heading of the tractor was in fact more stable in the simulation (see Figure \ref{fig:crabheading}). Most likely this is
due to testing condition differences such as terrain and slip values as well as the real
world tractor having covered more distance in its test than the digital twin.

\begin{figure}
   \centering
   \includegraphics[width=0.95 \linewidth]{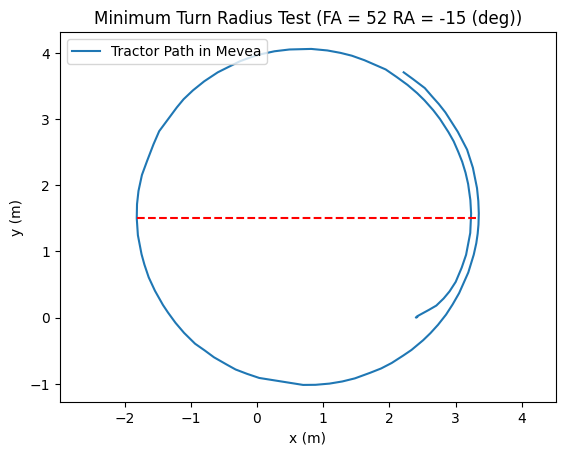}
   \caption{Minimum Turn Radius}
   \label{fig:min_turn_circle}
\end{figure}

The last test to evaluate the steering modeling was the minimum turn radius. Thus the test was carried out with these values. Data from the recorded CAN GNSS
messages containing latitude and longitude were then transformed back into a cartesian
x,y system for plotting (Figure \ref{fig:min_turn_circle}).

The Mevea simulation yielded a minimum turn radius of less than 6 meters. This is
roughly a meter less than what is reported in standard tractor's specification sheet. During the
simulation, it was clearly seen that the front wheels had a good amount of slip, correctly so, given the terrain they were on. Adversely, for a specification sheet, the turn radius
would have been provided for as little of slip as possible, thus explaining this discrepancy.

\begin{figure}
   \centering
   \includegraphics[width=0.95\linewidth]{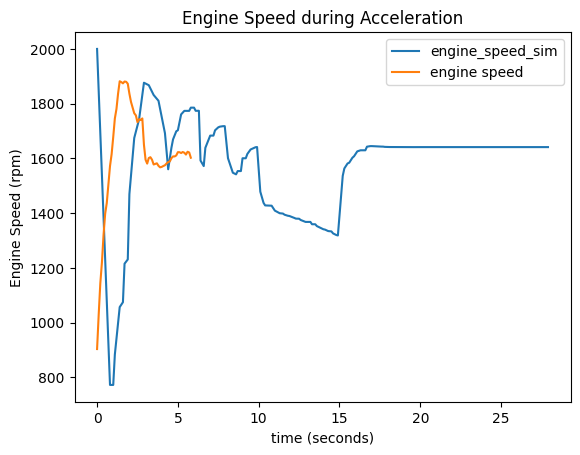}
   \caption{Engine Speed while Accelerating}
   \label{fig:enginespeed}
\end{figure}

\begin{figure}
   \centering
   \includegraphics[width=0.95\linewidth]{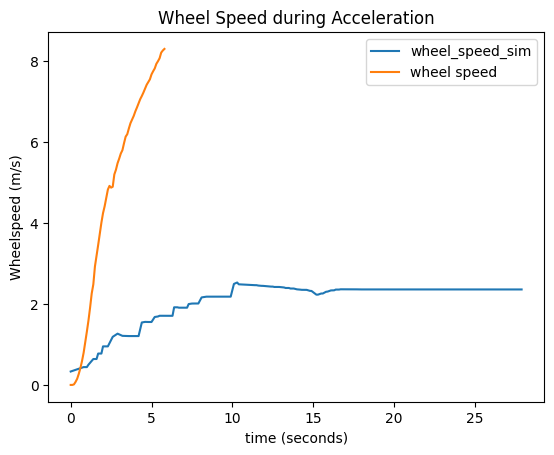}
   \caption{Wheel Speed while Accelerating}
   \label{fig:wheelspeed}
\end{figure}

As the acceleration test does not contain any 'set' inputs to try and hold still, percent
error over time here does not provide much information. Rather, simply viewing the
tractor engine (Figure \ref{fig:enginespeed}) and wheel speeds (Figure \ref{fig:wheelspeed}) during the acceleration phases allow for much more
insightful takeaways.

From these results, a couple of main ideas become clear. First, the difference between
the true power-split hydrostatic CVT transmission and the simple simulated step
transmission displays how much more efficient and effective the CVT is in accelerating
by keeping the rotational speed of the engine at its maximum torque range. Secondly,
the lack of available true mass and inertia data coupled with the unknown CVT
parameters play a crucial role in defining the tractor's agility, as one would expect. Of
course, other factors like ground friction values or tire pressure also contribute to these
differences, but they are much less impactful than what is shown here.

\FloatBarrier

\section{Summary and Conclusions}

Regarding the vehicle dynamics, it can be concluded that the steering mechanism of the
digital twin functions well. There are, however, several changes that need to be made to
the powertrain, should the necessary data become available. As for the CAN
implementation, the same code that is used to communicate within the real tractor has
been applied to the digital twin. Thus, while the data flow is slightly altered due to the
text files, the human interaction with the communication network remains very similar.
There are, however, no organizational processes such as the source address claiming
procedure of true CAN ISOBUS communication networks because there are no different
devices, just data flowing.

This project has provided a clear look into what tools are available for digital twin
modeling across all tasks. The exporting of 3D Farming Simulator 25 files makes it easy
to have the exact tractor in the simulation and several key data fields such as the engine
torque curve, weights, steering limitations or tire sizes are all quickly available from the
tractor specification sheet. However, more detailed parameters like inertia tensors for
each body, transmission values or tire frictions require significantly more effort. Mevea's front loader tutorial is therefore a very helpful tool in starting
to find ballpark values for some of these parameters.
With the transition of more new tractor transmissions to power split CVTs, a simpler
process for defining such in Mevea would be a valuable addition to the software. Additionally, while power-split CVT models already publicly exist, for
example in MATLAB, the problem remains that parameters specific to this transmission
are not fully available.
Ultimately, the goal of creating an initial digital twin model of an agricultural tractor
with a focus on CAN communication was achievable with the Mevea Simulation
Software. Basic research questions can now be addressed from any location in any
condition and the foundation for an improved digital twin has been laid.

As sensible next steps to this project, several topics present themselves.
First, for researchers looking to improve on this model, the biggest gain certainly appears
to be in the drivetrain and inertia values for the bodies. Without having the right
drivetrain implementation, the calculated rough inertia tensors did not allow for stable
testing. Similarly, assumptions such as solid bodies or semi-solid for axles had to be
made to get the mass properties in SolidWorks as exact mechanic model was not available. If the true
body properties can be found, this could provide a much more stable starting point for
transmission creation and testing. As mentioned early in the text, further consideration
of Mevea's compatibility with MATLAB Simulink can also provide for various strategies
to tackle the drivetrain.

On the CAN communication side, solutions to either decrease bus load or perhaps add capacity will help. Alternatively, other options besides text files can be investigated if cheaper virtual CAN channel solutions become available directly within Mevea similar to PCAN.
If instead, the focus is on adding new features to the model, Mevea’s capabilities in adding various sensors such as Lidar or including hydraulic functionalities is certainly interesting. Including a PTO to run implements would also be valuable to investigate dynamic behavior between the implement and tractor. However, this of course requires the modeling of an implement as well.
For purely visual improvements, manually changing the color of the various bodies in Blender will also give a quick boost to the simulation experience.
Finally, Mevea offers a wide range of output data, so other applications where a digital twin could be helpful are likely possible with their simulation software as well.

\begin{ack}
   We thank Mevea Oy and their employees for their support during this project.
   We thank Michael Maier for his help in performing the field tests with the real tractor.
\end{ack}

\bibliographystyle{IEEEtran} 
\bibliography{literature}             

\newpage
\onecolumn

\appendix
\section{Additional figures}    

\begin{figure}[ht]
   \centering

   \begin{subfigure}{0.48\columnwidth}
      \centering
      \includegraphics[width=\linewidth]{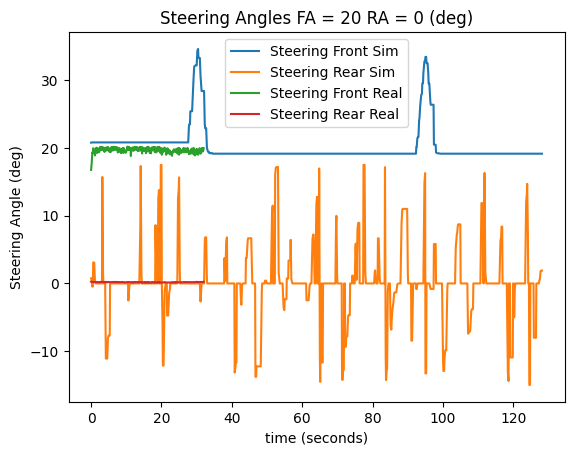}
      \label{fig:steer200}
      \caption{}
   \end{subfigure}
   \hfill
   \begin{subfigure}{0.48\columnwidth}
      \centering
      \includegraphics[width=\linewidth]{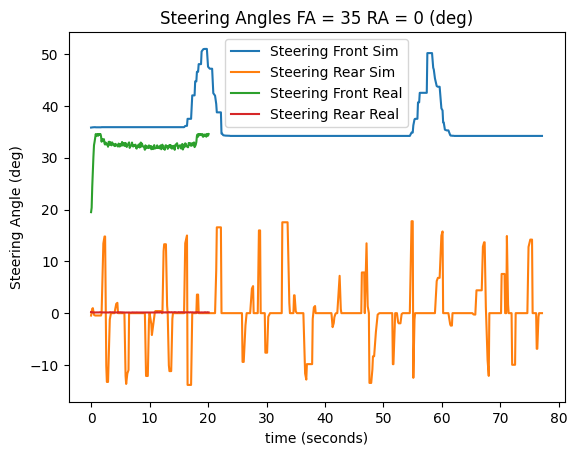}
      \label{fig:steer350}
      \caption{}
   \end{subfigure}
   \
   \begin{subfigure}{0.48\columnwidth}
      \centering
      \includegraphics[width=\linewidth]{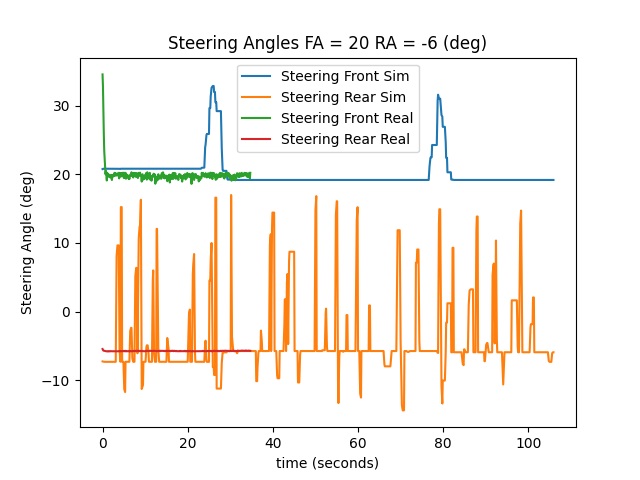}
      \label{fig:plot3}
      \caption{}
   \end{subfigure}
   \hfill
   \begin{subfigure}{0.48\columnwidth}
      \centering
      \includegraphics[width=\linewidth]{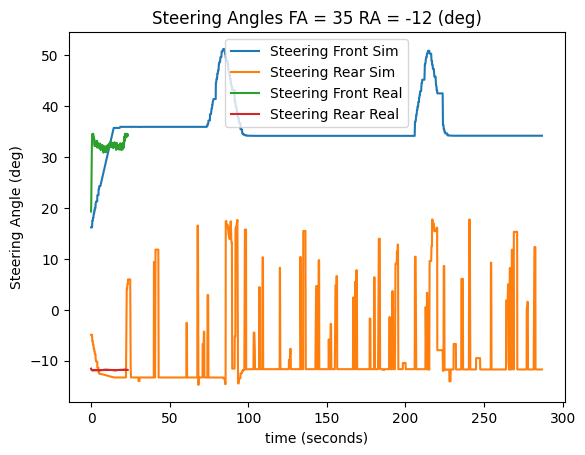}
      \label{fig:plot4}
      \caption{}
   \end{subfigure}
   \label{fig:plots_bottom1}
   \caption{Turning Steering Angles Over Testing Window}
   \label{fig:appendix1}
\end{figure}

\begin{figure}[ht]
   \centering

   \begin{subfigure}{0.48\columnwidth}
      \centering
      \includegraphics[width=\linewidth]{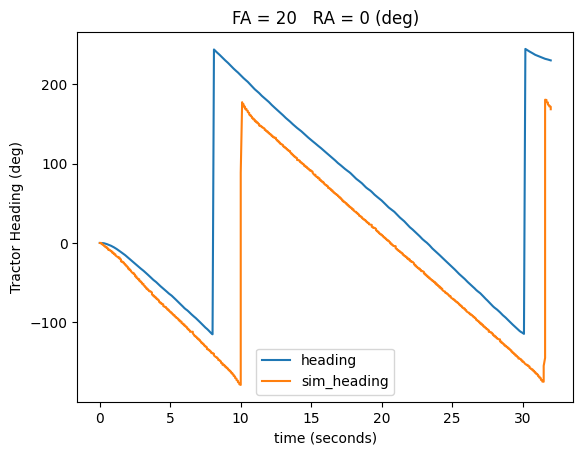}
      \label{fig:head200}
      \caption{}
   \end{subfigure}
   \hfill
   \begin{subfigure}{0.48\columnwidth}
      \centering
      \includegraphics[width=\linewidth]{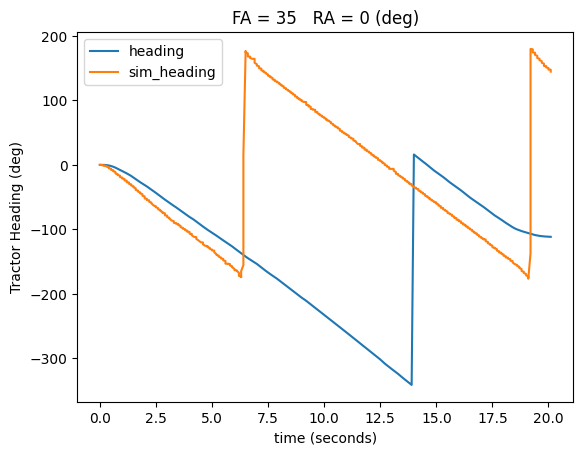}
      \label{fig:head350}
      \caption{}
   \end{subfigure}
   \
   \begin{subfigure}{0.48\columnwidth}
      \centering
      \includegraphics[width=\linewidth]{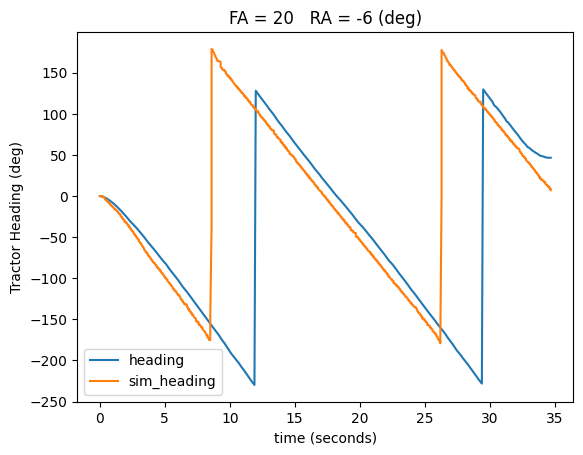}
      \label{fig:head20-6}
      \caption{}
   \end{subfigure}
   \hfill
   \begin{subfigure}{0.48\columnwidth}
      \centering
      \includegraphics[width=\linewidth]{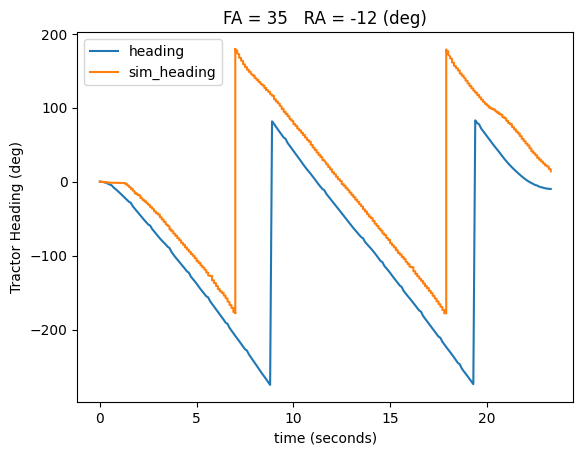}
      \label{fig:head35-12}
      \caption{}
   \end{subfigure}
   \label{fig:plots_bottom2}
   \caption{Heading Values while Turning}
   \label{fig:appendix2}
\end{figure}

\end{document}